\RequirePackage{rotating}
\documentclass[a4paper,fleqn,usenatbib]{mnras}

\usepackage{newtxtext,newtxmath}

\usepackage[T1]{fontenc}
\usepackage{ae,aecompl}
\usepackage{hhline}

\usepackage{graphicx}	
\usepackage{amsmath}	
\usepackage{caption}
\usepackage{rotating}

\newcommand\msun{\mbox{M$_{\odot}$}}

\newcommand\kms{\mbox{km~s$^{-1}$}}
\newcommand\HI{H\,{\sc i}}

\newcommand\sofia{{\sc sofia}}

\title[WALLABY Pre-Pilot Survey: UDGs]{WALLABY Pre-Pilot Survey: Ultra-Diffuse Galaxies in the Eridanus Supergroup}

\author[B.-Q. For et al.]{
  B.~-Q. For,$^{1,2}$\thanks{E-mail: biqing.for@uwa.edu.au}
  K. Spekkens,$^{3}$
  L. Staveley-Smith,$^{1,2}$
  K. Bekki,$^{1}$
  A. Karunakaran,$^{4}$
  B. Catinella,$^{1,2}$
  \newauthor
  B.~S. Koribalski,$^{5,6}$  
  K. Lee-Waddell,$^{1,7,8}$
  J.~P. Madrid,$^{9}$
  C. Murugeshan,$^{7,2}$ 
  J. Rhee,$^{1,2}$
  T. Westmeier,$^{1,2}$
  \newauthor
  O.~I. Wong,$^{7,1,2}$
  D. Zaritsky,$^{10}$
  R. Donnerstein$^{10}$\\
  $^{1}$International Centre for Radio Astronomy Research, University of Western Australia, 35 Stirling Hwy, Crawley, WA 6009, Australia\\
  $^{2}$ARC Centre of Excellence for All Sky Astrophysics in 3 Dimensions (ASTRO 3D), Australia \\
  $^{3}$Royal Military College of Canada, PO Box 17000, Station Forces, Kingston, ON K7K7B4, Canada\\
  $^{4}$Instituto de Astrof$\acute{i}$sica de Andaluc$\acute{i}$a, CSIC, Glorieta de la Astronom$\acute{i}$a, E-18080, Granada, Spain\\
  $^{5}$CSIRO Astronomy and Space Science, Australia Telescope National Facility, P.O. Box 76, NSW 1710, Australia\\
  $^{6}$School of Science, Western Sydney University, Locked Bag 1797, Penrith, NSW 2751, Australia\\
  $^{7}$CSIRO Space $\&$ Astronomy, PO Box 1130, Bentley, WA 6102, Australia\\
  $^{8}$International Centre for Radio Astronomy Research, Curtin University, Bentley, WA 6102, Australia \\
  $^{9}$Department of Physics and Astronomy, The University of Texas Rio Grande Valley, Brownsville TX 78520, USA\\
  $^{10}$Steward Observatory and Department of Astronomy, University of Arizona, 933 N. Cherry Ave., Tucson, AZ 85719, USA\\
}
\date{Accepted 2023 September 20. Received 2023 September 11; in original form 2023 March 30}
\pubyear{2023}

\begin{document}
\label{firstpage}
\pagerange{\pageref{firstpage}--\pageref{lastpage}}
\maketitle

\begin{abstract}
  We present a pilot study of the atomic neutral hydrogen gas (\HI) content of
  ultra-diffuse galaxy (UDG) candidates. 
  In this paper, we use the pre-pilot Eridanus field data from the Widefield ASKAP $L$-band Legacy
  All-sky Blind Survey (WALLABY) to search for \HI\ in UDG candidates found in
  the Systematically Measuring Ultra-diffuse Galaxies survey (SMUDGes).
  We narrow down to 78 SMUDGes UDG candidates within the maximum radial extents of the Eridanus subgroups for this study. 
  Most SMUDGes UDGs candidates in this study have effective radii smaller than 1.5 kpc and
  thus fail to meet the defining size threshold.
  We only find one \HI\ detection, which we classify as a low-surface-brightness dwarf. Six putative UDGs are \HI-free.
  We show the overall distribution of SMUDGes UDG candidates on the size-luminosity relation and compare them with low-mass dwarfs on the
  atomic gas fraction versus stellar mass scaling relation.
  There is no correlation between gas-richness and colour indicating that colour is not the sole
  parameter determining their \HI\ content. The evolutionary paths that drive galaxy morphological
  changes and UDG formation channels are likely the additional factors to affect the \HI\ content of putative UDGs.
  The actual numbers of UDGs for the Eridanus and NGC~1332 subgroups are
  consistent with the predicted abundance of UDGs and the halo virial mass relation, except for the NGC~1407 subgroup, which has
  a smaller number of UDGs than the predicted number.
  Different group environments suggest that these putative UDGs are likely formed via the satellite accretion scenario.

\end{abstract}




\section{Introduction}

Low-surface-brightness (LSB) galaxies have been studied for decades
(e.g., \citealp{Impey88, Bothun91, Dalcanton97}).
With the advancement of optical imaging instruments and search techniques, a large population of
extreme LSB galaxies has been uncovered (e.g., \citealp{AvD14, Z19}). Among them, the
discovery of tens to hundreds of spatially extended extreme LSB galaxies in the Coma cluster has 
reinvigorated the interest in studying these objects among observers and theorists \citep{vd15, Koda15}.
These so-called ultra-diffuse galaxies (UDGs) are typically defined to have an effective radius ($r_{\rm eff}$) $\geqslant$ 1.5~kpc
and a central $g$-band surface brightness ($\mu_{0, g}$) $\geqslant$ 24~mag~arcsec$^{-2}$ \citep{vd15}.
Given that
this definition is mostly motivated by observational constraints, some studies suggest that UDGs
may be a sub-class of the LSB dwarf population \citep{Conselice18, Habas20, Marleau21, LC20}. As they represent an  
extreme end of the LSB dwarf population, they are important in testing galaxy formation models \citep{BK12, Sawala16}. 

UDGs are prevalent across all environments. They have been found in clusters
(Coma cluster, e.g. \citealp{vd15, Koda15, Yagi16}; Virgo cluster, e.g. \citealp{Mihos15,Junais22};
Hydra cluster, e.g. \citealp{Iodice20}; and other clusters, see \citealp{Lee20} and references therein);
in galaxy groups (Hickson Compact groups, e.g. \citealp{RT17b, Shi17}; NGC 5485 group, \citealp{Merritt16})
and in the field (e.g. \citealp{Prole19, Prole21}). Their physical properties also vary across environments.
For example, they are generally red (quiescent), smooth and gas-poor in dense environments
but blue (star-forming), irregular and gas-rich in low-density environments \citep{RT17b,Kadowaki21}.
Their dark matter (DM) content has also sparked an intense debate about their
nature and formation mechanisms. 
Observational evidences suggest that
some of them are embedded in dwarf-sized DM halos \citep{BT16, C19}
and in more massive DM halos \citep{vd15, Z17, Forbes20}.
Some UDGs also exhibit peculiar properties,
such as high DM fractions \citep{Beasley16} and
an offset from the established baryonic Tully-Fisher relation \citep{MP20, K20}.
These pecularities challenge galaxy formation models. 
It is unclear if the offset from the baryonic Tully-Fisher relation is real or the result
of difficulties in measuring reliable inclinations, hence rotational velocities.  
It is also unclear how gas-rich blue UDGs form in low-density environments and how they relate to the cluster UDGs,
which tend to be gas-poor. 

Several hypotheses have been proposed to form UDGs. There are two main categories which
are driven by internal and external processes.

Internal processes:
\begin{itemize}
\item UDGs that are formed in dwarf sized halos might have higher than average spin parameters. The higher specific angular
momentum of the halo prevents gas from effectively collapsing into a dense structure, which explains
their extended size. In this scenario, field UDGs are predicted to be gas-rich \citep{Rong17, AL16}.  
\item In the NIHAO (Numerical Investigation of a Hundred Astrophysical Objects; \citealp{Wang15}) simulation field, 
UDGs can be formed via repeated star formation episodes during their early evolution, which
drives the gas out to larger radii. A non-negligible \HI\ gas mass of 10$^{7-9}$~\msun\ is predicted
for isolated field UDGs \citep{DC17}. It is worth noting that
UDGs may have lower star formation efficiencies than normal dwarfs despite being gas-rich \citep{KF22}.
\end{itemize}
External processes:
\begin{itemize}
\item UDGs may be failed $L_{\rm *}$ ($M_{\rm *}\sim10^{11}$~\msun) galaxies that do not form stars at the
rate expected for their halo mass due to star formation being quenched via ram-pressure or tidal effects \citep{vd15, YB15, Carleton21, Martin19, JS22}.
\item Present-day UDGs are formed from excess energy and angular momentum in merging low-mass galaxies early on ($z > 1$) \citep{Wright21}.
\item Strong tidal interactions with larger galaxies may also form diffuse tidal dwarf galaxies (TDGs) that are similar to UDGs \citep{Bennet18, Roman21}.
\item UDGs can be formed via tidal heating of normal dwarfs \citep{Jones21, Iodice21}. 
\end{itemize}
While these formation mechanisms are still under debate, 
a wide variety of properties may suggest that they are formed via a combination of the above proposed
mechanisms in different environments.

While deep optical imaging allows us to identify UDG candidates, 
one limitation is their distance determination.
This is in part the reason that UDG searches largely associate candidates to
clusters and groups at known distances.
While some distances to UDGs have been obtained, sample sizes remain
small due to a large amount of time that is required for spectroscopic follow-up observations 
(see e.g., \citealp{Kadowaki17, RL18, Kadowaki21, MN19, Emsellem19}). Observations of \HI-bearing 
UDGs allow an easier measurement of redshifts 
(inferring kinematic distances), which allows clear separation of foreground dwarfs from UDGs.
Such measurements also provide \HI\ masses and linewidth/rotation velocities  
for disentangling the formation mechanisms. 
Targeted \HI\ follow-up studies on nearby blue and star-forming UDGs have therefore been conducted and have shown to yield 
\HI\ masses consistent with the theoretical prediction \citep{Bellazzini17, Papastergis17, SK18, Scott21}. 
The untargeted Arecibo Legacy Fast ALFA (Arecibo $L$-band Feed Array) extragalactic \HI\ survey (ALFALFA; \citealp{G05}) data
have also proven to be useful in studying \HI-bearing UDGs in large numbers \citep{Leisman17, J19}.
Recently, an \HI\ pilot survey of
optical selected UDG candidates using the Robert C. Byrd Green Bank Telescope (GBT) has also been conducted \citep{K20}.
The ongoing and previous \HI\ studies are mostly utilising single-dish telescopes, 
which have better sensitivity than interferometers, albeit at the cost of lower angular resolution.

The Widefield ASKAP $L$-Band Legacy All-sky Blind SurveY (WALLABY, \citealp{Koribalski20})
makes use of the large field of view of the Australian Square Kilometre Array Pathfinder (ASKAP; \citealp{Johnston07, Hotan21})
to image \HI\ galaxies out to a redshift $z\sim0.1$ and to cover most of the southern hemisphere.
With the survey's high angular resolution of 30\arcsec\ and root-mean-square (RMS)
sensitivity of 1.6 mJy per beam per 18.5 kHz channel, 
WALLABY early science studies have been able to recover many gas-rich low-mass dwarf galaxies (see e.g. \citealp{For19, Kleiner19, For21}).
These galaxies were not resolved as individual sources in previous single-dish surveys, such as the 
\HI\ Parkes All-Sky Survey (HIPASS; \citealp{Barnes01}). WALLABY will be the first southern \HI\ survey to provide a large number of
\HI\ redshifts and physical parameters for those \HI-bearing UDGs that are identified as candidates in the DESI Legacy Imaging Surveys \citep{Dey19}.
This will allow us to investigate the proposed formation mechanisms from the \HI\ perspective across environments. 

\subsection{Eridanus Supergroup}

The concentration of galaxies in the region of Eridanus constellation was first noted by \citet{Baker33}. 
A later study by \citet{dv75} found that Group 31 and galaxies around the NGC 1332 and NGC 1209 formed the “Eridanus Cloud”.
This cloud lies on the Eridanus-Fornax-Dorado filamentary structure and is extended to the south and in front of the
“Great Wall” ($\sim$4000~\kms) \citep{costa88, Willmer89}.
Its structural complexity has drawn some debate regarding its nature.  
\citet{Willmer89} described it as a cluster made up of three or four subclumps. 
On the other hand, \citet{OD05a} considered that the galaxies in the region as loose groups
and form an intermediate evolutionary stage between the Ursa-Major group and the Fornax cluster.
\citet{B06} re-analysed this region using the 6dF Galaxy Survey (6dFGS; \citealp{Jones04}) and 
concluded that this region is occupied by 3 distinct groups, namely the NGC~1407 ($v=1658\pm26$~\kms), NGC~1332 ($v=1474\pm18$~\kms)
and Eridanus ($v=1638\pm5$~\kms) groups. 
These groups also form part of the supergroup, which is defined as a group of groups that may eventually merge to form a cluster.

The Eridanus supergroup is an interesting system as it is on the evolutionary path to cluster assembly similarly
to the Ursa-Major supergroup \citep{Wolfinger16}.
There are only a few known supergroups in the Universe that allow 
us to better understand galaxy evolutionary pathways \citep{Tran09}.
Galaxies in the Eridanus supergroup 
are more \HI-deficient as compared to galaxies in the Ursa-Major Supergroup and in the field \citep{For21, CM21}. 
There are two enormous \HI\ clouds in the Eridanus supergroup
without optical counterparts \citep{Wong21}, and the importance of tidal interactions in the
Eridanus supergroup has been recently demonstrated \citep{Wang22}. 

In this paper, we present a pilot study of \HI\ content of optically identified UDG candidates in the WALLABY pre-pilot Eridanus field and
discuss implications for their formation mechanisms.  
This paper is structured as follows. 
Section 2 describes the selection criteria of UDG candidates,
the methodology used to search for \HI\ and the derivation of their physical parameters. 
In Section 3, we perform the analysis of their distribution with respect to the low-mass dwarf population on
the atomic gas fraction vs stellar mass scaling relation, 
the predicted number of UDGs based on the virial masses of the host system, tidal or ram-pressure stripping as a
possible formation channel, 
gas-richness as compared to other UDGs in group and cluster environments. 
We summarise our findings and discuss future work in Section 4. 

Throughout the paper, we adopt a $\Lambda$ cold dark matter cosmology model ($\Lambda$CDM) with
$\Omega_{\rm M} = 0.27$, $\Omega_{\rm K} = 0$, $\Omega_{\rm \Lambda} = 0.73$ and $H_{\rm 0} = 73$ \kms\ Mpc$^{-1}$.
These are the default parameters for distances and cosmologically-corrected quantities in the 
NASA/IPAC Extragalactic (NED) database \citep{Spergel07}.

\section{UDG Candidates}

We select the UDG candidates from the third 
Systematically Measuring Ultra-diffuse Galaxies survey (SMUDGes) catalogue \citep{Z22}.
This catalogue focuses on identifying UDG candidates from
the southern portion of the ninth data release (DR9) of the DESI Legacy Imaging Surveys \citep{Dey19}. 
The classification is performed using a modified version of a deep learning model with visual confirmation. 
Objects with low surface brightness
($\mu_{\rm 0, g} \geqslant$ 24~mag arcsec$^{-2}$) and large angular extent (effective radii; $r_{\rm eff} \gtrsim$ 5.3\arcsec,   
which corresponds to $r_{\rm eff} \geq$ 2.5~kpc at the distance of the Coma cluster),
are set as the main selection criteria for the imaging data search in SMUDGes. 
We refer the reader to a detailed description of the image processing and the automated method for
identifying the UDG candidates in \citet{Z19, Z21}.

\begin{figure}
  \includegraphics[width=\columnwidth]{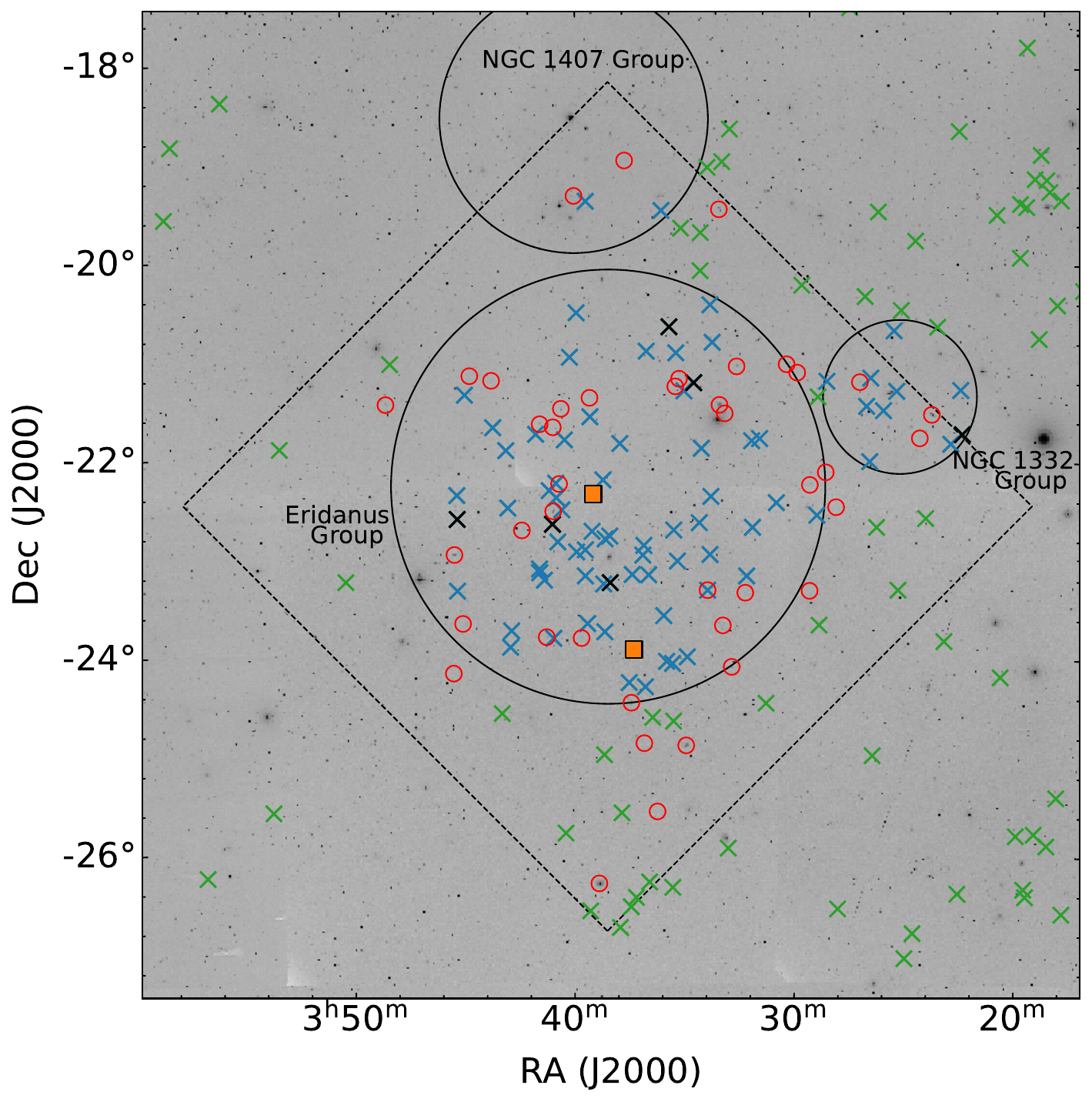}
  \caption{Ultra-diffuse galaxy candidates in the Eridanus field (crosses)
    from the SMUDGes 3rd catalogue overlaid onto an optical Digital Sky Survey (DSS-2) $r$-band image.
    The maximum radial extent of 
    Eridanus (0.8~Mpc), NGC~1332 (0.3~Mpc) and NGC~1407 (0.5~Mpc) are marked as black circles.
    UDG candidates within the maximum radial extents and selected for this study are shown as the blue crosses. Those UDG candidates
    that fall outside the maximum radial extent but within the Eridanus field are shown as the green crosses. 
    Those identified as putative UDGs ($r_{\rm eff}$ > 1.5~kpc at an assumed distance of 21~Mpc)
    are shown as the black crosses (see \S~2.2). 
    The WALLABY footprint is $\sim6\degr\times6\degr$, which is shown as the dashed diamond. 
    The red circles and orange squares represent the \HI\ detections and the \HI\ clouds in \citet{For21}, respectively.
    \label{onsky_udgs}}
\end{figure}

In Figure~\ref{onsky_udgs}, we show the distribution of SMUDGes
UDG candidates within a $10\degr\times10\degr$ area of the Eridanus
supergroup (crosses).
The maximum radial extent of the Eridanus, NGC~1407 and NGC~1332 groups \citep{B06} are represented as black circles.
The ASKAP observed area is marked with the dashed diamond, which fully covers the Eridanus group.
The red circles represent the \HI\ detections in \citet{For21}, hereafter F21.
The \HI\ sources in F21 generally have a higher
optical surface brightness than the SMUDGes UDG candidates. 
There are 97 and 78 SMUDGes UDG candidates within the WALLABY pre-pilot footprint and within the maximum radial extents of
the groups that comprise the Eridanus supergroup, respectively.

The completeness of the SMUDGes catalogue
is estimated to be a better representation of 
the population of large ($r_{\rm eff} > 2.5$~~kpc) UDGs beyond $cz \sim 1800$~\kms.
Given that the distance to the Eridanus supergroup is about 1/5 (one fifth) of the distance to Coma (100 Mpc), the vast majority 
of SMUDGes UDG candidates in this region are likely not
true UDGs.
A spatial variation of SMUDGes UDG candidates is seen in the Eridanus field.
Examining the observed footprints of DECaLS of this field, we find only minor differences in depth.
Excluding faint candidates does not change the spatial variation that we see in the field. 

\subsection{Searching for \HI}\label{gas}

To search for \HI\ in the SMUDGes UDG candidates, 
we extract a subcube at the position of each SMUDGes UDG candidate from the
WALLABY mosaicked cube of the Eridanus field. 
We extract the subcubes using three velocity ranges ($\sim$680--2500~\kms, $\sim$2000--7000~\kms\ and $\sim$7000--13000~\kms), 
which resulted in 97$\times$3 subcubes. Each subcube covers $0.1\degr\times0.1\degr$ in area.
The first velocity range covers the Eridanus supergroup.  

We run the Source Finding Application (\sofia\footnote{Available at \url{https://github.com/SoFiA-Admin/SoFiA-2/}};
\citealp{Serra15, Westmeier21}) to search for
\HI\ using 3.0, 3.5, 4.0, 4.5 and 5.0$\sigma$ detection thresholds. We note that 
sources presented in the F21 catalogue are detected with
a 5.0$\sigma$ threshold, where the local RMS is calculated from a larger area
and a wider velocity range than each extracted subcube.  
It is possible that changes of local RMS and lower thresholds might yield \HI\ detections 
that fall below the 5$\sigma$ detection threshold. 
Subsequently, we check the reliability plots from \sofia, and
all detections by eye using Hanning-smoothed cubes and DR9 DESI Legacy Imaging Survey images.
We only find one reliable \HI\ detection, which is also known as WALLABY~J033408$-$232125 $(cz$=1262~\kms) in the F21 catalogue.
This \HI\ detection is also a member of the Eridanus group (see F21). As a result of not finding any \HI\ detection
outside the maximum radial extents of these groups that make up the Eridanus supergroup (but still within WALLABY
observed footprints), we will focus on analysing
the 78 SMUDGes UDG candidates that are within the maximum radial extents of these groups for the rest of the paper. 
We note that there are two SMUDGes UDG candidates in the NGC 1332 group that fall outside of the ASKAP field of view.
The search algorithm of SMUDGes also does not detect any UDG candidates within 2\arcmin\ of the enormous \HI\ clouds
in the Eridanus group \citep{Wong21}. 

\subsection{Properties and Physical Parameters of SMUDGes UDG Candidates}

\begin{figure}
  \includegraphics[width=\columnwidth]{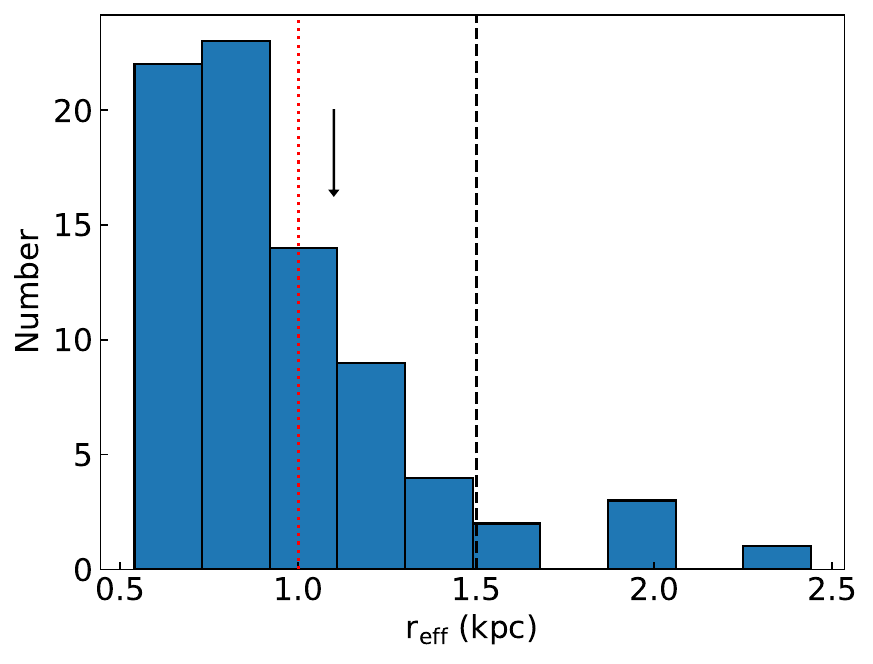}
  \includegraphics[width=\columnwidth]{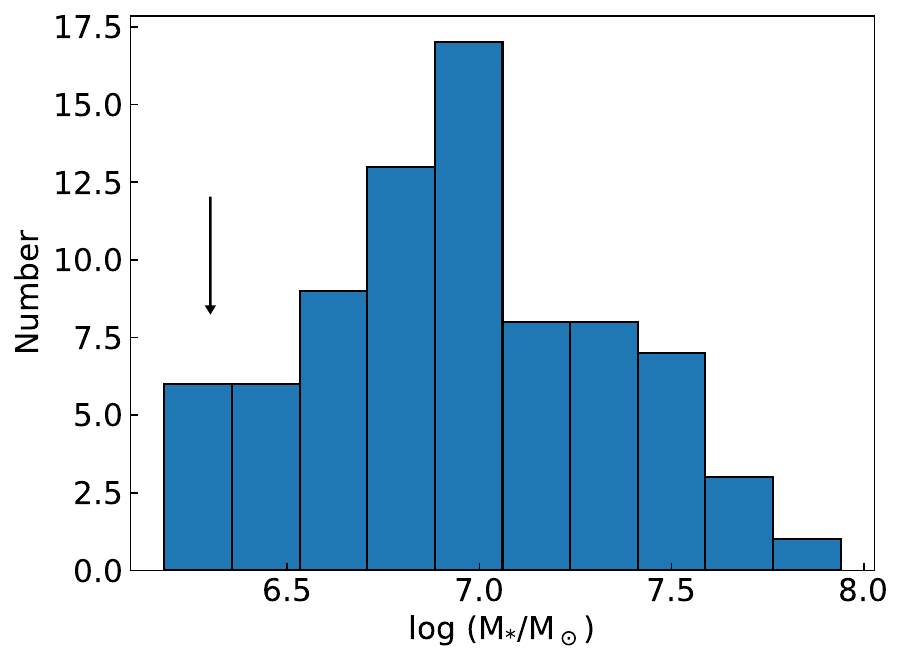}
    \includegraphics[width=\columnwidth]{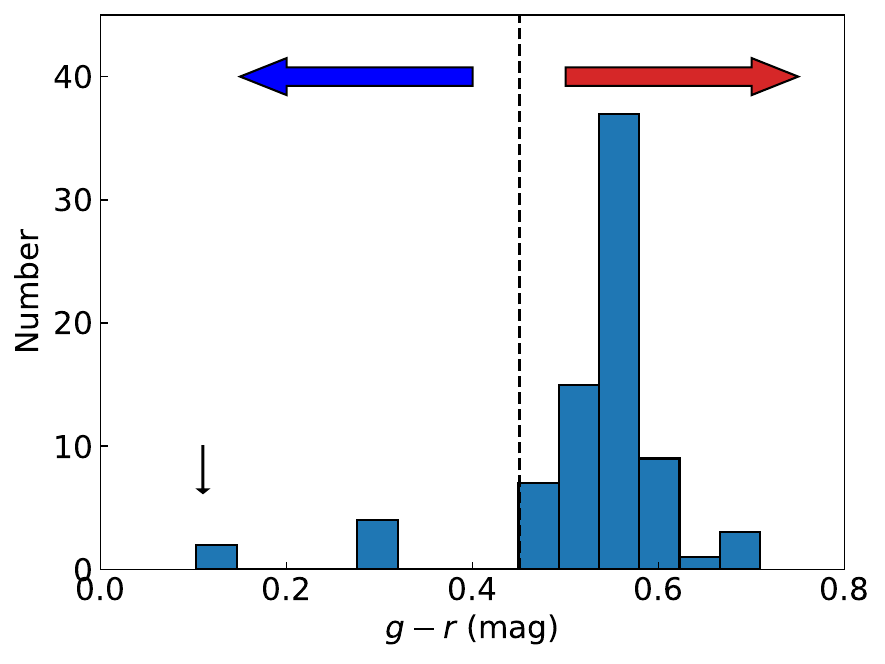}
    \caption{Histograms of $r_{\rm eff}$ in kpc at an assumed distance of 21~Mpc ({\it top}), $\log$~($M_{\rm *}/\msun$) ({\it middle}) and
      $g-r$ colour in magnitude ({\it bottom})
      for 78 SMUDGes UDG candidates that are within the maximum radial extents of the Eridanus subgroups.
      Bias corrections (if applicable) have been applied to $r_{\rm eff}$ and colour as given in \citet{Z22}.
      The black arrow indicates the position of the \HI\ detected source.
      {\it Top}: The dashed and dotted lines at 1.5~kpc and 1.0~kpc represent the boundaries of the defined observation and simulations UDG effective radius,
      respectively. 
      {\it Bottom}: The dashed line at $g-r =$ +0.45~mag represents the boundary of the defined red and blue colour for the SMUDGes UDG candidates.
      The definition is based on the joint distribution of colour ($g-r$) and S\'{e}rsic index ($n$) with the tail of red objects with $n$ > 1 and $g-r > +0.45$~mag \citep{Z21}. (see Table~\ref{tab1}). 
    \label{dist_hist}}
\end{figure}

Assuming that these 78 SMUDGes UDG candidates are group members of the Eridanus subgroups,
we adopt a luminosity distance of $D_{\rm L}$ = 21~Mpc (see Section~5 of F21) to calculate their $r_{\rm eff}$.
To obtain the stellar masses,
we employ the mass-to-light ratio ($M/L$) relation in \citet{Z09}\footnote{An initial mass function of \citet{Chabrier03} is adopted.}
as follows:

\begin{equation}
\log ({M_{*}/M_{\odot}}) = -0.804 + 1.654(g-r) + \log ({L_{r}/L_{\odot}}),
\end{equation}
where $g-r$ is an extinction-corrected colour and $L_{r}$ is the $r$-band luminosity measured from the DR9 DESI Legacy Survey
images. 
The absolute magnitude of the sun ($M_{\rm sun, abs}$) in different DES wavebands is given in \citet{Willmer18}. 
The $r$-band absolute magnitude is given as $M_{r, \rm abs} = r - 5\log(D_{\rm L}) + 5 - A_{\rm r}$,
where $D_{\rm L}$ is the luminosity distance in pc. We adopt the $A_{\rm r}$ (extinction)
value in the $r$-band
used for the SMUDGes catalogue \citep{Z22}.
In Figure~\ref{dist_hist}, we show the distributions of $r_{\rm eff}$, $M_{*}$ and $g-r$ for our 
SMUDGes UDG candidates. 

The definition of UDGs varies significantly in the literature. 
The widely accepted definition, i.e. $r_{\rm eff} \geqslant$ 1.5~kpc and $\mu_{0, g} \geqslant$ 24~mag arcsec$^{-2}$,
stems from the samples in \citet{vd15}.
There is a wide range of parameters being explored as selection criteria by various studies. 
For example, \citet {Yagi16} and \citet{RT17b} use $r_{\rm eff} > 0.7$~kpc and $r_{\rm eff} > 1.3$~kpc as their minimum radius
definition, respectively. 
This is solely motivated by observational constraints rather than a physical reason.   
There are other constraints which have variously been suggested, 
such as stellar mass,
absolute magnitude and/or luminosity, to
explicitly limit UDGs to dwarf mass populations (see \citealp{Mihos15,Iodice20,Lim20}).

\begin{figure}
  \includegraphics[width=\columnwidth]{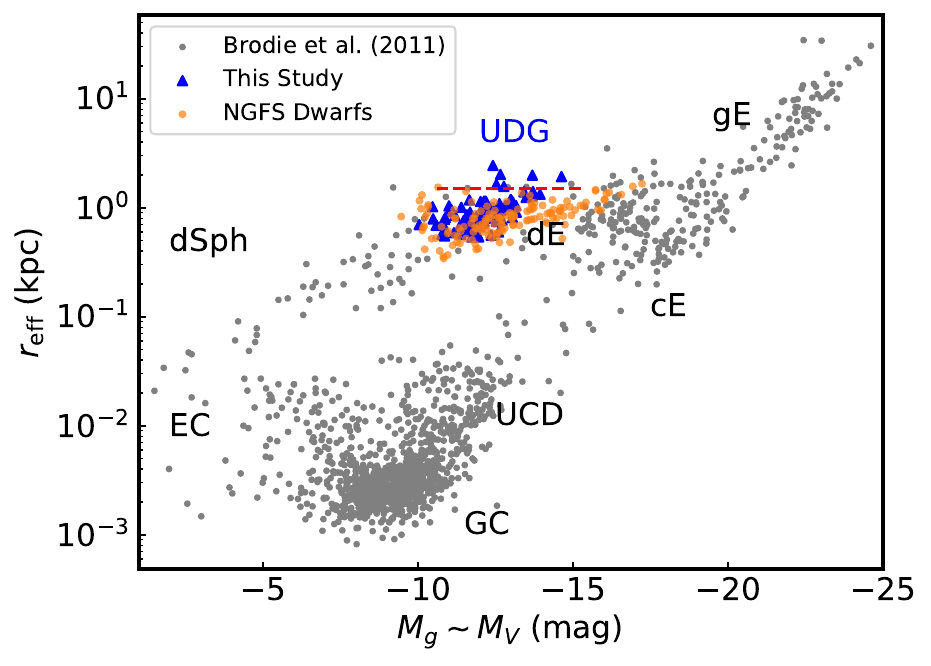}
  \caption{Size--luminosity relation of dwarf galaxy populations.
    Grey dots, blue triangles and orange dots represent objects compiled by \citet{Brodie11}, this study and NGFS dwarfs \citep{Eigenthaler18}, respectively.
    Size is represented by effective radius ($r_{\rm eff}$) in kpc with SMUDGes UDG candidates (LSB dwarfs + putative UDGs) at an assumed distance of 21~Mpc. The red dashed line indicates $r_{\rm eff}$ = 1.5~kpc.
    Absolute magnitudes in  the DES $g$-band (extinction corrected) are used for our SMUDGes UDG candidates (LSB dwarfs + putative UDGs) sample,
    which is equivalent to the absolute $V$-band magnitude in \citet{Brodie11}.
    dSph: dwarf Spheroidal; dE: dwarf Elliptical; gE: giant Elliptical; cE: compact Elliptical; UCD: Ultra Compact Dwarf;
    GC: Globular Cluster; EC: very faint, Extended Cluster; UDG: Ultra-Diffuse Galaxy. \label{pop}}
  \end{figure}

In this paper, we consider putative UDGs to have $r_{\rm eff}$ > 1.5~kpc. This allows us to make a direct
comparison with previous studies.
With this definition, we obtain six putative UDGs among our 78 SMUDGes UDG candidates,
with five and one belonging to the Eridanus and NGC~1332 groups, respectively.
WALLABY~J033408$-$232125 has $r_{\rm eff}$ > 1.1~kpc and
hence will be considered as an LSB dwarf rather than a UDG in this work.\footnote{Note that if we were to consider the definition of
$r_{\rm eff} > 1.0$~kpc from the theoretical NIHAO simulation,
the total number of putative UDGs would increase to 22 and 
WALLABY~J033408$-$232125 would be considered as a UDG instead based on its redshift.}
In Figure~\ref{pop}, we overplot our SMUDGes UDG candidates (LSB dwarfs + putative UDGs) onto
the size--luminosity relation of dwarf galaxy populations (grey dots) compiled
by \citet{Brodie11}\footnote{\url{https://sages.ucolick.org/spectral_database.html}} and
the central dwarf galaxy population in the Next Generation Fornax Survey (NGFS; \citealp{Eigenthaler18}).
\citet{Brodie11} samples are limited to objects that have confirmed distances either by
spectroscopy, resolved stellar populations or surface brightness fluctuations.  
Our SMUDGes UDG candidates (LSB dwarfs + putative UDGs) sample extends to the fainter end of the dwarf Spheroidal (dSph) galaxies.
The NGFS dwarf galaxy population (nucleated and non-nucleated) overlaps with the UDGs parameter space and is
extended toward the brighter end of the dwarf Elliptical (dE). 
This has drawn some debate if UDGs are at the large end of the dwarf locus or are branching out into its own sequence. 
They are generally low-mass ($M_{\rm *} < 10^{8}$~\msun) and the majority
of them are fairly red in colour ($g-r > 0.45$ as defined in \citealp{Z21}).
The median S\'{e}rsic index ($n$) is $\sim0.8$ in the SMUDGes catalogue.
This value is consistent with theoretical predictions \citep{Jiang19}. 

For non-\HI\ detections, we calculate their upper \HI\ mass limit ($z\sim0$) as follows:

\begin{equation}
M_{\rm HI, lim} = 236\times10^{3} \times S_{\rm int} \times D_{\rm L}^{2},
  \end{equation}
where $S_{\rm int}$ = 5$\sigma \times\Delta v$ is the \HI\ integrated flux, in Jy~\kms\ and $D_{\rm L}$ is the luminosity distance of
21~Mpc.
We adopt a fiducial velocity width ($\Delta v$) of 30~\kms\ (see e.g., \citealp{Jones21}). 
1$\sigma$ noise level is calculated from 100 source free channels of each subcube that covers the velocity range of 680--2500~\kms.
We summarise the properties and physical parameters of our SMUDGes UDG candidates (LSB dwarfs + putative UDGs) sample
in Table~\ref{tab1}. 

\begin{table*}
  \centering
  \begin{minipage}{175mm}
  \caption{Properties and derived parameters of SMUDGes UDG candidates (LSB dwarfs + putative UDGs).}\label{tab1}
\begin{tabular}{llcccccrrrl}
\hline
ID	&	Designation	&	$\alpha$ (J2000)	&	$\delta$ (J2000)	&	$g-r$	&	$M_{\rm r}$	&	$\log M_{*}$	&	$\log M_{\rm HI}$	&	$\log (M_{\rm HI} / M_{*})$	&	$R_{\rm eff}$	&	$R_{\rm eff}$	\\
	&		&	(deg)	&	(deg)	&	(mag)	&	(mag)	&	(M$_{\odot}$)	&	(M$_{\odot}$)	&		&	(arcsec)	&	(kpc)	\\
(1)	&	(2)	&	(3)	&	(4)	&	(5)	&	(6)	&	(7)	&	(8)	&	(9)	&	(10)	&	(11)	\\
\hline
\multicolumn{11}{c}{Eridanus Group} \\                                                                                                                                                                                       
\hline    
1	&	SMDG~0336515$-$242027	&	54.21476	&	$-$24.34079	&	0.555	&	$-$13.723	&	7.45	&	< 7.55	&	< 0.11	&	8.94	&	0.91	\\
2	&	SMDG~0337349$-$241754	&	54.39532	&	$-$24.29840	&	0.501	&	$-$12.693	&	6.95	&	< 7.56	&	< 0.62	&	11.30	&	1.15	\\
3	&	SMDG~0335396$-$240610	&	53.91490	&	$-$24.10278	&	0.504	&	$-$11.899	&	6.63	&	< 7.55	&	< 0.92	&	9.93	&	1.01	\\
4	&	SMDG~0335557$-$240456	&	53.98221	&	$-$24.08214	&	0.592	&	$-$13.230	&	7.31	&	< 7.55	&	< 0.24	&	7.48	&	0.76	\\
5	&	SMDG~0335006$-$240205	&	53.75237	&	$-$24.03474	&	0.555	&	$-$13.378	&	7.31	&	< 7.55	&	< 0.24	&	10.34	&	1.05	\\
6	&	SMDG~0342509$-$235621	&	55.71210	&	$-$23.93930	&	0.577	&	$-$11.349	&	6.53	&	< 7.66	&	< 1.13	&	5.58	&	0.57	\\
7	&	SMDG~0340559$-$235101	&	55.23296	&	$-$23.85031	&	0.519	&	$-$12.394	&	6.86	&	< 7.67	&	< 0.81	&	7.96	&	0.81	\\
8	&	SMDG~0338400$-$234705	&	54.66684	&	$-$23.78466	&	0.562	&	$-$13.298	&	7.29	&	< 7.64	&	< 0.36	&	9.65	&	0.98	\\
9	&	SMDG~0342478$-$234626	&	55.69903	&	$-$23.77390	&	0.570	&	$-$12.843	&	7.12	&	< 7.66	&	< 0.54	&	9.99	&	1.02	\\
10	&	SMDG~0339260$-$234204	&	54.85834	&	$-$23.70116	&	0.569	&	$-$12.147	&	6.84	&	< 7.66	&	< 0.82	&	8.47	&	0.86	\\
11	&	SMDG~0336039$-$233707	&	54.01629	&	$-$23.61862	&	0.549	&	$-$12.674	&	7.02	&	< 7.58	&	< 0.56	&	7.32	&	0.75	\\
12	&	SMDG~0345106$-$232201	&	56.29428	&	$-$23.36689	&	0.536	&	$-$12.292	&	6.84	&	< 7.63	&	< 0.79	&	8.11	&	0.83	\\
13	&	SMDG~0334081$-$232128$^{\ddagger}$&	53.53354&	$-$23.35785	&	0.103	&	$-$12.733	&	6.30	&	  7.54	&	  1.24	&	10.72	&	1.09	\\
14	&	SMDG~0338435$-$231802	&	54.68114	&	$-$23.30051	&	0.555	&	$-$14.235	&	7.65	&	< 7.63	&	< $-$0.02	&	12.95	&	1.32	\\
15	&	SMDG~0338261$-$231711$^{\dagger}$&	54.60877&	$-$23.28646	&	0.466	&	$-$12.999	&	7.01	&	< 7.64	&	< 0.63	&	16.07	&	1.64	\\
16	&	SMDG~0341202$-$231539	&	55.33436	&	$-$23.26092	&	0.592	&	$-$13.117	&	7.27	&	< 7.63	&	< 0.36	&	7.35	&	0.75	\\
17	&	SMDG~0339319$-$231306	&	54.88274	&	$-$23.21826	&	0.509	&	$-$11.983	&	6.68	&	< 7.63	&	< 0.96	&	7.43	&	0.76	\\
18	&	SMDG~0332252$-$231233	&	53.10507	&	$-$23.20915	&	0.548	&	$-$14.487	&	7.74	&	< 7.54	&	< $-$0.20	&	13.06	&	1.33	\\
19	&	SMDG~0336444$-$231222	&	54.18518	&	$-$23.20610	&	0.538	&	$-$12.559	&	6.95	&	< 7.62	&	< 0.66	&	11.20	&	1.14	\\
20	&	SMDG~0337279$-$231213	&	54.36616	&	$-$23.20352	&	0.554	&	$-$12.238	&	6.85	&	< 7.63	&	< 0.78	&	8.90	&	0.91	\\
21	&	SMDG~0341326$-$231108	&	55.38587	&	$-$23.18560	&	0.537$^{*}$&	$-$12.198$^{*}$	&	6.81	&	< 7.62	&	< 0.81	&	11.60	&	1.18	\\
22	&	SMDG~0341331$-$230852	&	55.38812	&	$-$23.14772	&	0.462	&	$-$11.061	&	6.23	&	< 7.61	&	< 1.38	&	6.74	&	0.69	\\
23	&	SMDG~0335286$-$230353	&	53.86915	&	$-$23.06482	&	0.597$^{*}$&	$-$11.070$^{*}$	&	6.46	&	< 7.58	&	< 1.13	&	9.99	&	1.02	\\
\hline
\end{tabular}
    {\it Note.} This table is available in its entirety as Supporting Information with the electronic version of the paper.
    A portion is shown here for guidance regarding its form and content. 
    $\dagger$: Putative UDG. $\ddagger$: Also known as WALLABY~J033408$-$232125.
    $*$: No bias correction due to being flagged as inaccurate \citep{Z22}.
    $\#$: Within maximum radial extent of the NGC 1332 group but outside the WALLABY observed footprints. 
  Col (1): Identification number.
  Cols (2)--(4): Designation, $\alpha$ and $\delta$ (J2000) coordinates are based on the SMUDGes catalogue \citep{Z22}. 
  Col (5): Extinction and bias (if applicable) corrected $g-r$ magnitude.
  Col (6): Absolute magnitude in $r$-band.
  Col (7): Stellar mass in logarithmic scale.
  Col (8): Upper limit of \HI\ mass, with the exception of WALLABY~J033408$-$232125, in logarithmic scale.
  Col (9): Atomic gas fraction.  
  Col (10): Effective radius in arcsec, which is derived using DR9 DESI Legacy Survey stacked images of $grz$-bands. Bias correction has been applied if applicable. 
  Col (11): Effective radius in kpc for an assumed distance of 21~Mpc.
\end{minipage}
\end{table*}

\section{Analysis}

\subsection{Atomic Gas Fraction-Stellar Mass Scaling Relation}

The atomic gas fraction ($M_{\rm HI}/M_{*}$) versus stellar mass ($M_{*}$) scaling relation 
allows us to understand the physical processes that regulate the conversion of gas
into stars and drive the changes in galaxy morphology (see e.g., \citealp{J17, SC22}).

The Extended $GALEX$ Arecibo SDSS Survey (xGASS; \citealp{Catinella18}) investigates
this relation for a representative sample of $\sim$1200 galaxies,
selected from the SDSS DR7 \citep{A09} by stellar mass and redshift only
(9.0 < $\log$~($M_{*}/\msun$) < 11.5
and 0.01 < $z$ < 0.05),
and observed down to a gas fraction limit of 2-10\%,
depending on $M_{*}$. Its stellar mass selected \HI\
sample shows a clear linear relation of increasing $\log$~($M_{\rm HI}/M_{*}$)
with decreasing $\log$~($M_{*}/\msun$),
but it is unclear if the trend continues to rise below the $\log$~($M_{*}/\msun$) = 9.0 limit of the survey.
Using the ALFALFA.40 sample with stellar masses derived from SDSS spectra and photometry, 
\citet{Maddox15} show that the gas fraction follows the
same trend as in the xGASS sample at higher stellar mass ($M_{*} > 10^{9}$~\msun) but flattens out
at the lower
stellar mass end indicating a higher gas content in the low-mass regime. 

To further investigate the trend in the low-mass regime, F21 compare various low-mass dwarf 
samples with the \citet{Maddox15} empirical relations (see Figure~12 of F21). It is evident that 
the sub-sample of gas-rich Local Volume dwarfs selected
from ALFALFA.40 in \citet{Huang12} does not show the flattening trend as seen in \citet{Maddox15}.
In Huang et al's sub-sample, the atomic gas fraction continues to rise with decreasing $M_{*}$. This is due to 
stellar masses in the ALFALFA.40 sample being  
underestimated for low-mass galaxies, which is a known issue with the SDSS reduction pipeline \citep{Huang12}.
The derived stellar masses via 
the spectral energy distribution (SED) fitting method in \citet{Huang12} are higher by comparison. 
F21 show that the sample from the Survey of \HI\ in Extremely Low-mass Dwarfs (SHIELD; \citealp{McQuinn21}) also
supports a non-flattening trend in the low-mass regime. The SHIELD sample mostly consists of isolated dwarf galaxies. 
It is unclear if the low-mass dwarf population
in the Eridanus supergroup follows such a trend due to a lack of low-mass ($< 10^{8}$~\msun) galaxies in F21.

\begin{figure}
  \includegraphics[width=\columnwidth]{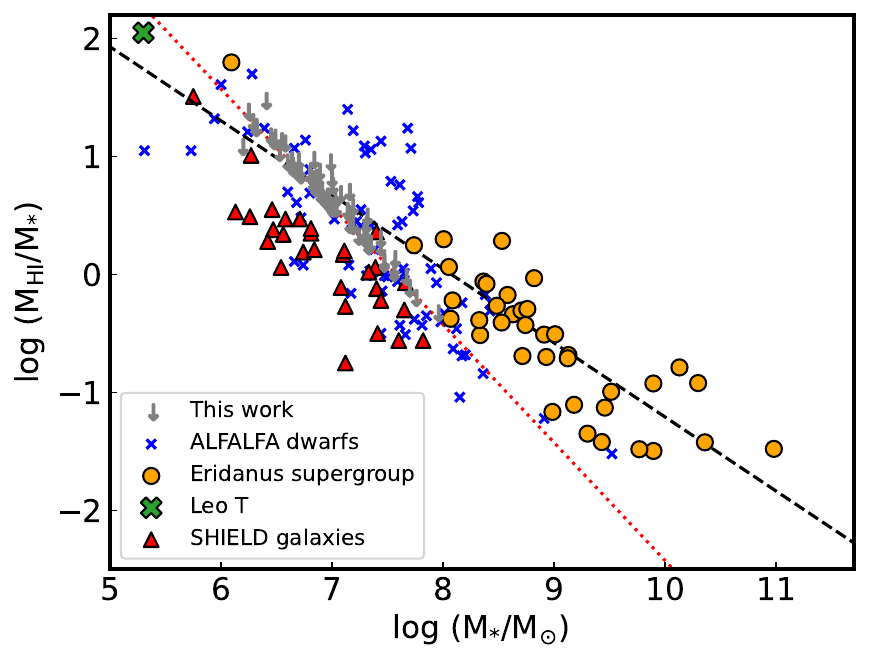}
  \caption{Atomic gas fraction ($M_{\rm HI}$/$M_{*}$) versus stellar mass ($M_{*}$) scaling relation in logarithmic scale.
    SHIELD galaxies (red triangles), dwarf galaxies from the ALFALFA.40 subsample (blue crosses),
    the Eridanus supergroup with \HI\ detections (orange dots), SMUDGES UDG candidates (LSB dwarfs + putative UDGs)
    (grey arrows) with distances assumed to be equal to that of
    the Eridanus supergroup (non-\HI\ detections) and
    Leo T (green cross) are plotted for comparison. The black dashed line is derived in F21. The red dotted line
    is the gas-fraction sensitivity limit (5$\sigma$) at the distance of the Eridanus supergroup (21~Mpc).
    1$\sigma$ is equivalent to 2.4~mJy (F21). 
    \label{gasfrac}}
  \end{figure}

With a large number of low-mass galaxies in this study, we revisit the atomic gas fraction 
scaling relation in the low-mass regime of the Eridanus supergroup.
In Figure~\ref{gasfrac}, we show the distribution of the ALFALFA.40 sub-sample (blue crosses; \citealp{Huang12}),
the SHIELD sample (red triangles; \citealp{McQuinn21}), the Eridanus supergroup sample (orange circles; F21),
the Leo T dwarf galaxy (green cross; \citealp{AO18}) and our SMUDGes UDG candidates (LSB dwarfs + putative UDGs) sample (grey points)
on the atomic gas fraction scaling relation. The atomic gas fraction values for our SMUDGes UDG candidates (LSB dwarfs + putative UDGs)
sample are set to be
their upper limit with the exception of one \HI\ detected LSB dwarf (see section~\ref{gas}).
The dashed line is adopted from F21 and is for guidance only.
Our sample falls along the WALLABY sensitivity limit line, which also lies within
the scatter of the gas-rich dwarf population. 
We note that galaxies in the Eridanus supergroup are generally
more \HI-deficient as compared to galaxies in other galaxy groups (see e.g. \citealp{For19}).
Nevertheless, the finding of a non-flattening trend based on ALFALFA.40 sub-sample and SHIELD sample
in the low-mass regime suggests that low-mass, high gas fraction galaxies might be rarer than expected.
It is inconclusive regarding the flattening trend in the low-mass regime using our sample. 
It would be useful to revisit such relation in the low-mass regime with the full WALLABY survey in the future. 

\subsection{Gas Richness}

\begin{figure}
  \includegraphics[width=\columnwidth]{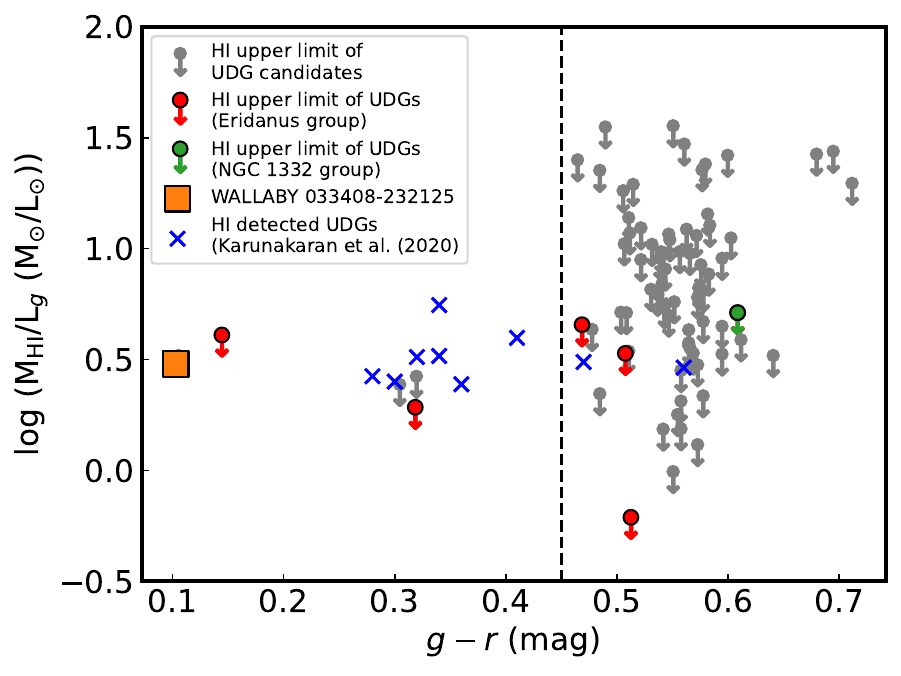}
    \includegraphics[width=\columnwidth]{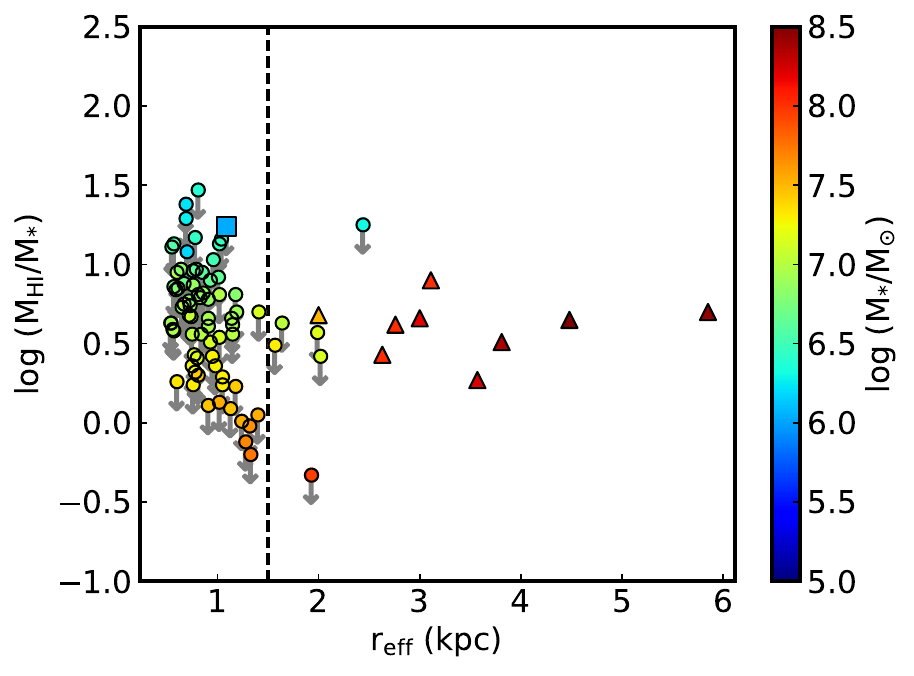}
    \caption{Top: Distribution of $\log$($M_{\rm HI}/L_{g}$ (\msun/$L_{\rm \odot}$)) vs $g-r$.
      All magnitudes are extinction and bias corrected.
      The dashed line divides the red and blue populations at $g-r$ = 0.45.
      The orange square and blue crosses represent the \HI\ detected LSB dwarf in our study 
      and \HI\ detected UDGs in \citet{K20}, respectively.
      The arrows show the upper limit of $M_{\rm HI}/L_{g}$ for  
      five and one putative UDGs in the Eridanus and NGC 1332 groups are shown as red and green arrows, respectively.  
      Bottom: Distribution of $\log (M_{\rm HI}/M_{\rm *})$ vs $r_{\rm eff}$.
      Bias corrections have been applied to $r_{\rm eff}$ values.
      The colorscale indicates the stellar masses. The SMUDGes UDG candidates (LSB dwarfs + putative UDGs) sample
      in our study are represented by circles (\HI\ non-detections) and square (\HI\ detection).
      \HI\ detected UDGs in the Coma cluster are shown as triangles \citep{K20}. The arrows show the upper limit of $M_{\rm HI}/M_{\rm *}$.
    \label{gas_richness}}
  \end{figure}

We investigate the gas richness of our SMUDGes UDG candidates (LSB dwarfs + putative UDGs) sample using distance independent measurements. 
In Figure~\ref{gas_richness} (top panel), we show the distribution of
$\log$~($M_{\rm HI}/L_{g}$) versus $g-r$. The \HI\ upper limits of our sample are represented by the arrows, 
with six putative UDGs in red and green for the Eridanus and NGC~1332 groups, respectively.
The \HI\ detected UDGs in the Coma cluster (blue crosses; \citealp{K20}) and our \HI\ detected LSB dwarf (orange square) are also shown. 
The dashed line represents $g-r = 0.45$ that marks the blue and red color boundary. 
According to \citet{K20}, \HI\ detected UDGs are bluer and have more irregular morphologies, while the non-\HI\ detected
UDGs are redder and smoother in morphologies. We find that the two putative UDGs (red arrows) and
two LSB dwarfs (grey arrows) with $g-r < 0.45$
have a smooth morphology. While the sample is small, it is possible that different morphologies resulting from
different evolutionary paths might affect the gas content of the UDGs.
The majority ($\sim94\%$) of putative UDGs and LSB dwarfs 
in this study are red in colour. This is not a surprise given that UDGs in denser environments tend to be redder \citep{Kadowaki21}. 
The unusually blue colour of LSB dwarf, SMDG~0334081$-$232128 (also known as WALLABY~J033408$-$232125), suggests
ongoing star formation with a star formation rate estimated to be 0.0002~\msun~yr$^{-1}$ (F21).  

The bottom panel of Figure~\ref{gas_richness} shows the distribution of
atomic gas fraction $\log$~($M_{\rm HI}/M_{\rm *}$) versus $r_{\rm eff}$ with the colour scale representing stellar mass.
The triangles, square and circles represent the \HI\ detected UDGs in the Coma cluster \citep{K20},
LSB dwarfs with and without \HI\ in our study, respectively. 
A scatter of $\sim$1~dex is present for $r_{\rm eff} < 1.0$~kpc.  
There is no obvious trend beyond $r_{\rm eff} > 1.5$~kpc but our sample size is small.
UDGs in the Coma cluster are massive by comparison to our putative UDGs. 
The atomic gas fraction of Coma cluster UDGs also has $\log (M_{\rm HI}/M_{\rm *}$) < 1.0
indicating that the relative gas content is lower
than our sample. 

\subsection{Number of UDGs and Virial Masses of their Host Halos}

The number of UDGs ($N_{\rm UDG}$) is known to have a power-law relation with the virial masses ($M_{\rm 200}$)
of the host halos, $N_{\rm UDG} \propto$ $M^{\alpha}_{\rm 200}$ \citep{vdB16, vdB17, Janssens17, MP18, Lee20}.
The power-law index, $\alpha$, gives an indication of how effectively a galaxy is formed and survives its environment.
If $\alpha = 1$, the number of galaxies is directly proportional to the mass of the host halo. In this case, these galaxies
are not strongly affected by environmental effects. If $\alpha<1$, galaxies in low-density environments have
relatively higher number densities per unit mass of their host halos.
These galaxies are preferentially formed and survive in low-density environments (field or galaxy group).
If $\alpha>1$, these galaxies are formed more efficiently or survive longer in high-density environments (cluster) \citep{Lee20}. 

A list of UDG numbers and virial masses of galaxy groups and clusters has been compiled by \citet{Lee20} 
based on the same selection criteria as in this study.
\citet{Lee20} fit the $N_{\rm UDG} \propto$ $M^{\alpha}_{M_{\rm 200}}$ relation by
considering the data points with $M_{\rm 200} > 10^{13}$ \msun, $M_{\rm 200} > 10^{12}$ \msun\ and full range of $M_{\rm 200}$. 
There are fewer data points for fitting the relation on host halos with $M_{\rm 200} < 10^{13}$ \msun\
as there are fewer UDGs in low-mass halos. Using all available data in \citet{Lee20}, we obtain
a Pearson correlation coefficient of 0.94 indicating a tight correlation. 
Overall, their derived $\alpha$ is close to 1, which suggests that the formation and survival of UDGs are less
affected by the environment.  
However, we caution that the power-law fitting is affected by selection bias and small number statistics, especially
for lower mass host halos. This is clearly demonstrated by using the GAMA group sample, 
where \citet{vdB17} obtain $\alpha = 1.11$. 
Their sample is also less abundant per unit halo mass than the Hickson Compact Groups (HCGs; \citealp{RT17b}).
The deviation might be due to the lack of UDGs in the GAMA groups and the group properties, as loose galaxy groups
are not included in the HCGs.

Recently, \cite{KZ23} study the abundance of UDGs around 75 nearby Milky Way-like systems using literature satellite galaxy catalogues.
Their investigation bolsters the low halo mass end of the UDG abundance relation and finds a slope of $\alpha = 0.89$ for this relation.
Crucially, they demonstrate that there are various systematics (e.g. UDG definitions, photometric completeness, and host redshifts)
between various UDG abundance studies in the literature that can affect the result slopes and highlight the need for more uniform studies of this trend.
Nevertheless, as explained in their work, the majority of existing slopes hover around a slope of unity and
imply little to no effect of the environment on UDG abundance.

\begin{figure}
  \includegraphics[width=\columnwidth]{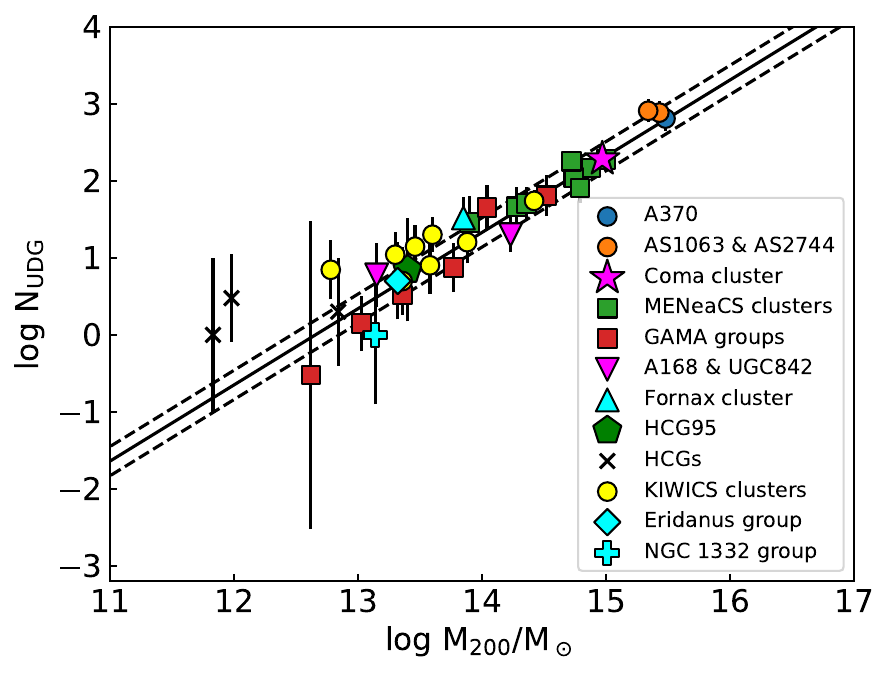}
  \caption{Number of UDGs and virial mass of their host system in logarithmic scale.
    The number of UDGs in other galaxy groups and clusters is compiled and listed in Table~7 of \citet{Lee20}.
    Blue circle: A370 \citep{Lee20};
    orange circles: AS1063 and AS2744 \citep{Lee17};
    magenta star: Coma cluster \citep{Yagi16};
    green squares: MENeaCS clusters \citep{vdB16};
    red squares: GAMA groups \citep{vdB17};
    magenta nablas: A168 and UGC842 \citep{RT17a};
    cyan triangle: Fornax cluster \citep{Venhola17};
    green pentagon: HCG95 \citep{Shi17};
    black crosses: HCGs \citep{RT17b};
    yellow circles: KIWICS clusters \citep{MP18};
    cyan diamond: Eridanus group (this study);
    cyan plus: NGC 1332 group (this study).
    The solid and dashed lines represent $\log$ ($N_{\rm UDG}$) $= 0.99 \times \log M_{\rm 200}/\msun -12.53\pm0.67$ and RMS of 0.19~dex. 
    \label{n_udg}}
  \end{figure}

To predict the number of UDGs in the Eridanus supergroup, we apply the power-law relation (with $\alpha$=0.99)
in \citet{Lee20} to the halo masses of the NGC 1407 ($7.9\times10^{13}$~\msun), NGC 1332 ($1.4\times10^{13}$~\msun)
and Eridanus ($2.1\times10^{13}$~\msun) groups as listed in \citet{B06}.
We obtain $N_{\rm UDG}$ = 3, 5 and 17 for the NGC~1332, Eridanus and NGC 1407 groups, respectively.
Our sample yields $1^{+3}_{-1}$, $5^{+8}_{-5}$ and $0^{+2}$ putative UDGs for the NGC~1332, Eridanus and NGC~1407 groups, respectively.
The uncertainties are based on the Poisson statistics. 
Taking into account the uncertainties in halo masses, the predicted numbers of UDGs for
the NGC~1332 and Eridanus groups are consistent with our finding (assuming the projected distance of 21~Mpc).
However, the number of UDGs in the NGC 1407 group is lower than the number predicted by the power-law relation ($N_{\rm UDG} = 17\pm6$).
This is not a surprise given that the number of SMUDGes UDG candidates
in the NGC 1407 group is small to begin with (refer to Figure~\ref{onsky_udgs}).
In Figure~\ref{n_udg}, we show the relation between $N_{\rm UDG}$ and $M_{\rm 200}$
for a sample of groups and clusters, including the new data. The fit suggests $\alpha$~=~0.99.

\begin{figure}
  \includegraphics[width=\columnwidth]{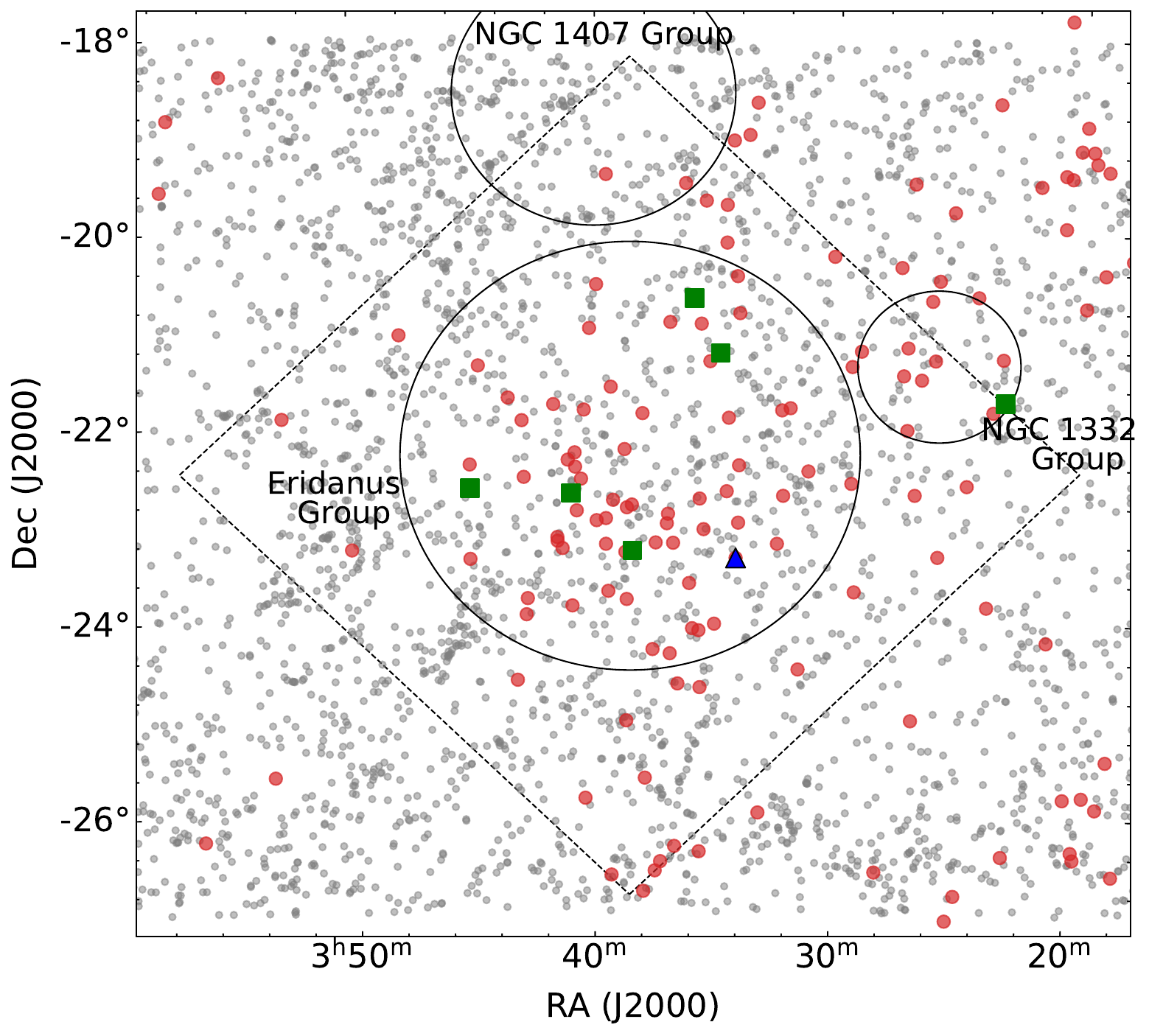}
  \caption{Spatial distribution of galaxies in the region of the Eridanus supergroup.
    Galaxies from the 2MASS All-Sky Extended Catalogue are shown as the grey dots.
    Putative UDGs and \HI\ detected LSB dwarf 
    are shown as the 
    green squares and the blue triangle, respectively.
    The red dots represent the SMUDGes UDG candidates in the Eridanus field.
    The black circles mark the maximum radial extents of each subgroup.
    The WALLABY footprint is $\sim6\degr\times6\degr$, which is shown as the dashed diamond. 
    \label{spatial_dist}}
  \end{figure}

To investigate if spatial density might contribute to a higher number of
SMUDGes UDG candidates in different groups, we plot the spatial projected
distribution of galaxies in the region of the Eridanus supergroup (grey dots) in Figure~\ref{spatial_dist}. 
These galaxies are extracted from the 2MASS All-Sky Extended Catalogue (XSC; \citealp{Jarrett00}).
Our putative UDGs and \HI\ detected LSB dwarf are represented by green squares and a blue triangle, respectively.
The non-\HI\ detected LSB dwarfs and other SMUDGes UDG candidates in the Eridanus field are represented by red dots. 
We find that the density of LSB dwarfs (initially as SMUDGes UDG candidates) in the Eridanus field
is at its highest in the Eridanus group as compared to the NGC~1407 and NGC~1332 groups.
The projected locations of putative UDGs do not show a correlation with the density of LSB dwarfs. 
The slightly different environments between these groups could explain the actual number of UDGs in the groups.

The NGC~1407 group has a centroid located 16~kpc from the large elliptical galaxy NGC~1407 \citep{B06}. It
is the only group in the supergroup that contains X-ray emission. Due to its  
high mass-to-light ratio, low spiral fraction and symmetric intragroup X-ray emission, it
is considered to be virialised. The higher X-ray luminosity of the NGC~1407 group compared to
other galaxy groups \citep{Miles04} also suggests that the NGC~1407 group is dynamically stable.
While NGC~1407 group is not a cluster, the presence of X-ray emission and a
large fraction of early-type galaxies suggest that
its evolutionary stage is more consistent with galaxy clusters (where UDGs are gas-poor) than young galaxy groups. 
On the contrary, the Eridanus group is not centred on any particular galaxy. Its centroid is 300~kpc from
the brightest elliptical galaxy (NGC~1395) in the group. 
It is a loose and dynamically young group that is yet to reach a stable stage of evolution (F21). 

The centroid of the NGC~1332 group is 43~kpc from the brightest lenticular galaxy (NGC~1332).
It lacks X-ray emission because it is not hot enough in a low-density environment and
as a virialised low-mass group. 
It is also not as dynamically mature as the NGC~1407 group \citep{B06}.
The formation of UDGs is possibly ongoing in the NGC~1332 and Eridanus groups. 
If satellite accretion is one of the formation
mechanisms for UDGs to exist in the Eridanus supergroup, it might explain why the mature group has fewer UDG candidates
to begin with as they might have been disrupted or merged into the central galaxy.
Further simulations might shed some light on the number of UDGs in various evolutionary stages of galaxy group.
There are also only a few foreground galaxies (based on spectroscopic redshift) within the Eridanus field.
  Their large projected distances with our putative UDGs would make them unlikely to be the host galaxy.  

\subsection{Tidal or Ram-Pressure Stripping}

Comparison of UDGs in the NIHAO simulations and in the simulation of a galaxy group \citep{Jiang19}
shows that satellite UDGs are mostly quiescent and gas-poor. 
In this scenario, satellite UDGs are presumably puffed up in the field and later quenched when falling
into a dense environment. The main quenching mechanisms in galaxy groups are tidal and/or ram-pressure
stripping. If the tidal mechanism is dominant,
the gas will be stripped and the stars 
will be removed from the outskirts of the satellite UDGs, which reduces their $r_{\rm eff}$ and $M_{\rm *}$.
We would also expect $M_{\rm *}$ to be lower for UDGs closer to the group centre due to stronger tidal effects.
The studies of UDGs in HCGs \citep{RT17b} and in the Coma cluster \citep{Alabi18}
support this scenario, with red UDGs predominantly 
located closer to the group and cluster centres.

In general, the NIHAO simulation is consistent with the observations of UDGs being
quiescent (red) in the inner part of the galaxy groups and 
star-forming (bluer) toward the outskirts of the galaxy groups.
However, simulations do not show an obvious radial gradient in size or stellar mass,
which suggests that tidal stripping might not be the dominant mechanism in quenching the UDGs in groups \citep{Jiang19}.
Following the evolutionary paths of satellite UDGs in simulations, \citet{Jiang19} show that 
tidally induced puffing is only partially responsible for the lack of radial extent, and the change in stellar mass
is small. 
While the sample is small, our study with the five putative UDGs in the Eridanus group
neither show colour nor stellar mass vs projected distance correlations. 
There are no such correlations for the rest of our LSB dwarfs either, which 
supports the scenario of UDGs being formed via accretion in the Eridanus and NGC 1332 groups.

\begin{figure}
  \includegraphics[width=\columnwidth]{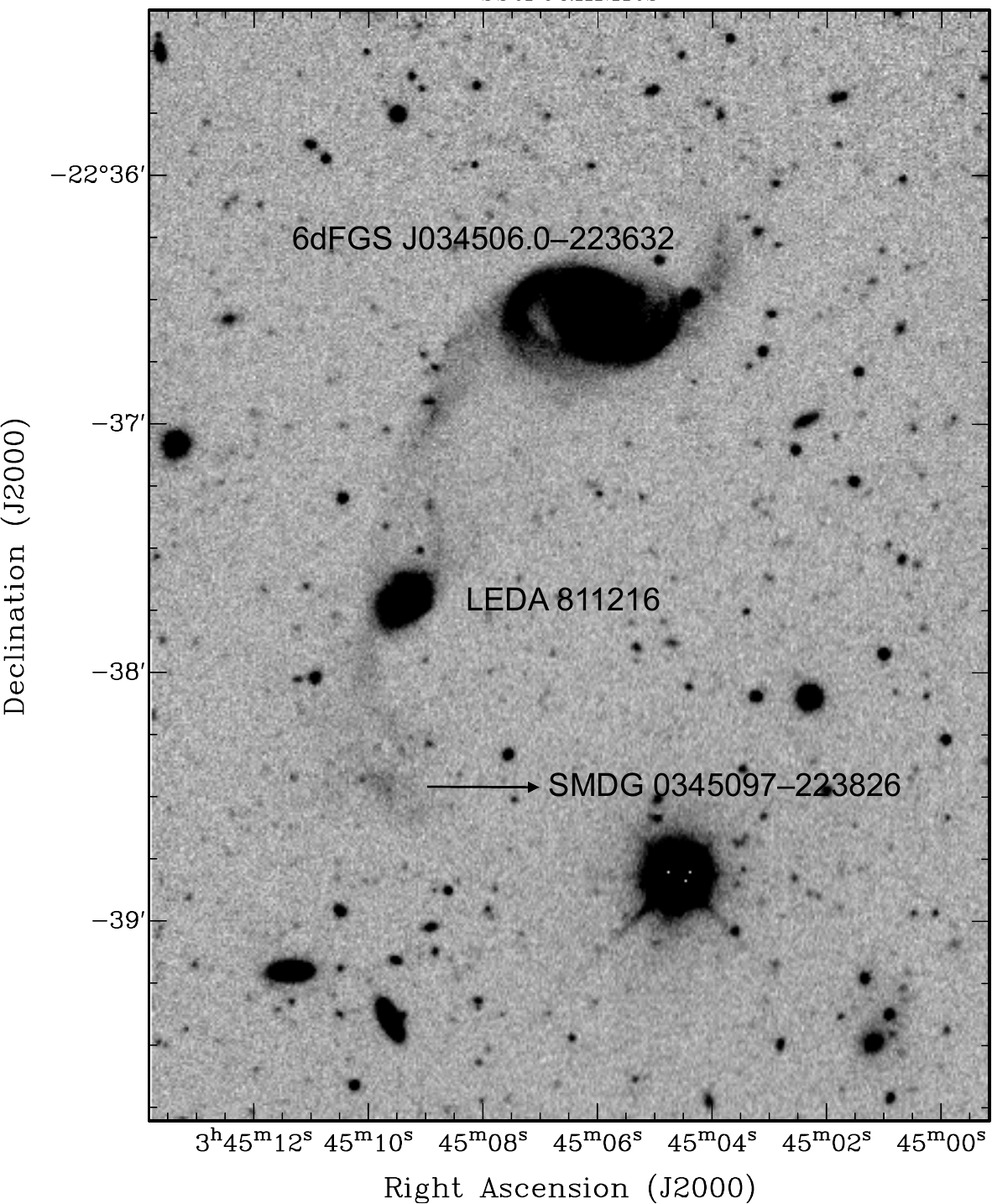}
  \caption{Co-added $g$-band image from the DR9 imaging Legacy Surveys.
    The locations of 6dFGS~J034506.0$-$22363, LEDA~811216 and SMDG~0345097$-$223826 are labelled. 
    \label{sstream}}
  \end{figure}

To investigate if our putative UDGs have experienced tidal heating, tidal or ram-pressure stripping origin, 
we search for any bright galaxies within 2\arcmin\ radius around them.
While UDGs with tidal origin are generally located close (< 20~kpc projected distance) to their parent galaxies (see e.g., \citealp{Jones21}),
they potentially form as far as 40~kpc in projected distance from their progenitors (see e.g., \citealp{Iodice20}).
We find that SMDG~0345097$-$223826 
appears to be located at the tail end of a stellar stream of 6dFGS~J034506.0$-$223632 (see Figure~\ref{sstream}).
There is also a fairly blue spiral galaxy, LEDA~811216, that is located south of 6dFGS~J034506.0$-$223632 and alongside the stellar stream.
The location of LEDA~811216 could be a projection effect or it could in fact be interacting with 6dFGS~J034506.0$-$22363.
We retrieve the spectroscopic and photometric redshifts of $0.04249\pm0.00015$ \citep{Jones09} and $0.010\pm0.006$ \citep{Zhou21} for 6dFGS~J034506.0$-$22363 and LEDA~811216, respectively.

The photometric redshift error is built upon with a few assumptions and using the random forest regression routine of a machine learning library, Scikit-Learn \citep{Pedregosa11}. 
It is subject to the training sets and does not include incompleteness in the training data or uncertainties in morphological parameters.
Luminous red galaxies (LRGs) are the main sample in the study of \citet{Zhou21}, which is biased toward higher redshift and redder objects.
LEDA~811216 is relatively blue with $g-r = 0.34$. Hence, the quoted photometric redshift of LEDA~811216 is unlikely to be accurate.

The slight disruption on the optical morphology of LEDA~811216 suggests that this pair is interacting.
We do not find any \HI\ in any of our velocity ranges at their locations. 
If SMDG~0345097$-$223826 is associated with 6dFGS~J034506.0$-$223632, we can rule out that it is a member of the Eridanus group.
At that redshift distance, SMDG~0345097$-$223826 would be considered to be large in size ($r_{\rm eff}$ > 13~kpc). 
We can not verify the nature and origin of SMDG~0345097$-$223826 without follow-up spectroscopic redshift observation.


\section{Summary and Future Work}

We use the WALLABY pre-pilot survey data of the Eridanus field (F21) to search for \HI\ in optically-identified UDG candidates
in the third release of the SMUDGes catalogue \citep{Z22}.
There are 78 UDG candidates within the maximum radial extents of the Eridanus subgroups and there is only one reliable \HI\ detection.
The detection is a confirmed member of the Eridanus group (see F21).
We investigate the properties and derive the physical parameters of these SMUDGes UDG candidates. 
Using $r_{\rm eff} > 1.5$~kpc and $\mu_{0, g}$ $\geqslant$ 24~mag~arcsec$^{-2}$ as the definition of a UDG,
we obtain six putative UDGs.
The rest is classified as LSB dwarfs.
We also find that 
our SMUDGes UDG candidates (LSB dwarfs + putative UDGs) sample is generally low-mass ($M_{\rm *} < 10^8$~\msun)
and most of them are fairly red in colour ($g-r > 0.45$).

It is inconclusive if our SMUDGes UDG candidates (LSB dwarfs + putative UDGs) sample yields a flattening trend at the low-mass regime while
examining the $M_{\rm HI}/M_{\rm *}$ versus $M_{\rm *}$ scaling relation. 
The distribution of gas richness versus colour shows no correlation in our SMUDGes UDG candidates (LSB dwarfs + putative UDGs) sample.
The two putative UDGs and two LSB dwarfs that have $g-r < 0.45$ appear to have smoother optical morphology.
There is no \HI\ detection among them. 
This supports the finding of \citet{K20}, which states that the optical morphology is
also an important parameter when looking for \HI\ in UDGs. 

We adopt the derived power-law relation of UDG number and the virial mass of their host halo in \citet{Lee20} to obtain
the predicted numbers of UDGs for the Eridanus subgroups. \citet{Lee20} predict
3, 5 and 17 for the NGC~1332, Eridanus and NGC~1407 groups, respectively.
The corresponding numbers of putative UDGs are $1^{+3}_{-1}$, $5^{+8}_{-5}$ and $0^{+2}$.
The lack of putative UDGs in the NGC~1407 group is likely
due to the evolutionary stage of that group. 
We investigate if tidal or ram-pressure stripping is a possible formation mechanism for UDGs
in the Eridanus supergroup. We find a putative UDG (SMDG~0345097$-$223826) that could have been formed via tidal heating/interaction as it
is located at the tail end of a stellar stream. 
If that was the case, this putative UDG would likely be a background UDG and would not be associated with the Eridanus group. 

It is known that the gas content of galaxies in the group environment is more diverse as it depends on the
evolutionary stages of the group. To understand if such variation is also observed for UDGs in groups, we
conduct a pilot study of \HI\ content of SMUDGes UDG candidates in the Eridanus supergroup using the
WALLABY data in conjunction with deep optical images and catalogue.
To extend what to expect for the WALLABY full survey, we cross-match the full WALLABY survey area
with the third release of the SMUDGes catalogue \citep{Z22}. We find $\sim$1750 SMUDGes UDG candidates within the 
overlapping survey area. With one \HI\ detection out of 78 SMUDGes UDG candidates, we could expect $\sim$22 \HI\ detections as
the lower limit for the full WALLABY survey. There are WALLABY survey areas that are currently not
covered by the DR9 DESI Legacy Survey, which is the third release of the SMUDGes catalogue based on.
In addition, we expect the \HI\ detection rate to be higher in the isolated and loose group environments.
The \HI\ redshift of isolate detections would be of great benefit as these are difficult to associate
with local overdensities.
In the future, this study will be expanded by using WALLABY current released pilot and full survey data,
which will allow us to probe the 
\HI\ content and UDG formation channels across environments.




\section*{Acknowledgements}

This research was supported by the Australian Research Council Centre of Excellence for All Sky Astrophysics in 3 Dimensions (ASTRO 3D),
through project number CE170100013.
This scientific work uses data obtained from Inyarrimanha Ilgari Bundara / the Murchison Radio-astronomy Observatory.
We acknowledge the Wajarri Yamaji People as the Traditional Owners and native title holders of the Observatory site.
CSIRO’s ASKAP radio telescope is part of the Australia Telescope National Facility (\url{https://ror.org/05qajvd42}).
Operation of ASKAP is funded by the Australian Government with support from the National Collaborative Research Infrastructure Strategy.
ASKAP uses the resources of the Pawsey Supercomputing Research Centre. Establishment of ASKAP, Inyarrimanha Ilgari Bundara,
the CSIRO Murchison Radio-astronomy Observatory and the Pawsey Supercomputing Research Centre are initiatives of the Australian Government,
with support from the Government of Western Australia and the Science and Industry Endowment Fund. \\

This research has made use of images of the Legacy Surveys. The Legacy Surveys consist of three individual and
complementary projects: the Dark Energy Camera Legacy Survey (DECaLS; Proposal ID \#2014B-0404; PIs: David Schlegel and Arjun Dey),
the Beijing-Arizona Sky Survey (BASS; NOAO Prop. ID \#2015A-0801; PIs: Zhou Xu and Xiaohui Fan), and the Mayall z-band Legacy Survey
(MzLS; Prop. ID \#2016A-0453; PI: Arjun Dey). DECaLS, BASS and MzLS together include data obtained, respectively, at the Blanco telescope,
Cerro Tololo Inter-American Observatory, NSF’s NOIRLab; the Bok telescope, Steward Observatory,
University of Arizona; and the Mayall telescope, Kitt Peak National Observatory, NOIRLab.
The Legacy Surveys project is honored to be permitted to conduct astronomical research on Iolkam Du’ag (Kitt Peak),
a mountain with particular significance to the Tohono O’odham Nation. \\

We thank the anonymous referee for their constructive comments to improve this manuscript.
BQF thanks A.~Bosma, A.~Boselli, B.~Holwerda, A.~L\'{o}pez-S\'{a}nchez, K.~McQuinn, G.~Muerer, J. Rom\'{a}n and P.~Zuo for their comments on the manuscript. 
KS acknowledges support from the Natural Sciences and Engineering Research Council of Canada (NSERC).
AK acknowledges financial support from the grant CEX2021-001131-S funded by MCIN/AEI/ 10.13039/501100011033
and from the grant POSTDOC\_21\_00845 funded by the Economic Transformation,
Industry, Knowledge and Universities Council of the Regional Government of Andalusia and financial support from the grant
PID2021-123930OB-C21 funded by MCIN/AEI/ 10.13039/501100011033, by “ERDF A way of making Europe” and by the "European Union".
DZ and RD gratefully acknowledge financial support for SMUDGes from NSF AST-1713841 and AST-2006785.
DZ thanks the Astronomy Department at Columbia University for their gracious welcome during his sabbatical.


\section*{Data availability}
The data underlying this article are available in the
article and in its online supplementary material. The processed
ASKAP data can be retrieved via CSIRO ASKAP Science Data Archive (CASDA)
with a given scheduling block identification number.
The DOI for the Eridanus data is \url{https://dx.doi.org/10.25919/0yc5-f769}.




\bibliographystyle{mnras}
\bibliography{ref} 


\section*{Supporting information}

Additional Supporting Information can be found in the online version
of this article.

Table~\ref{tab1}: Properties and derived parameters of selected SMUDGes sample. 





\bsp	
\label{lastpage}
\end{document}